# MGA TRAJECTORY PLANNING WITH AN ACO-INSPIRED ALGORITHM

Matteo Ceriotti

*Advanced Space Concepts Laboratory*

*Department of Mechanical Engineering, University of Strathclyde*

*75 Montrose Street, James Weir Building*

*Glasgow G1 1XJ, United Kingdom*

matteo.ceriotti@strath.ac.uk

Phone: +44 (0)141 548 5726

(Corresponding author)

Massimiliano Vasile

*Space Advanced Research Team*

*Department of Aerospace Engineering, University of Glasgow*

*James Watt South Building*

*Glasgow G12 8QQ, United Kingdom*

m.vasile@eng.gla.ac.uk

Phone: +44 (0)141 330 6465

## ABSTRACT

Given a set of celestial bodies, the problem of finding an optimal sequence of swing-bys, deep space manoeuvres (DSM) and transfer arcs connecting the elements of the set is combinatorial in nature. The number of possible paths grows exponentially with the number of celestial bodies. Therefore, the design of an optimal multiple gravity assist (MGA) trajectory is a NP-hard mixed combinatorial-continuous problem. Its automated solution would greatly improve the design of future space missions, allowing the assessment of a large number of alternative mission options in a short time. This work proposes to formulate the complete automated design of a multiple gravity assist trajectory as an autonomous planning and scheduling problem. The resulting scheduled plan will provide the optimal planetary sequence and a good estimation of the set of associated optimal trajectories. The trajectory model consists of a sequence of celestial bodies connected by two-dimensional transfer arcs containing one DSM. For each transfer arc, the position of the planet and the spacecraft, at the time of arrival, are matched by varying the pericentre of the preceding swing-by, or the magnitude of the launch excess velocity, for the first arc. For each departure date, this model generates a full tree of possible transfers from the departure to the destination planet. Each leaf of the tree represents a planetary encounter and a possible way to reach that planet. An algorithm inspired by Ant Colony Optimization (ACO) is devised to explore the space of possible plans. The ants explore the tree from departure to destination adding one node at the time: every time an ant is at a node, a probability function is used to select a feasible direction. This approach to automatic trajectory planning is applied to the design of optimal transfers to Saturn and among the Galilean moons of Jupiter. Solutions are compared to those found through more traditional genetic-algorithm techniques.

Multiple gravity assist, Interplanetary trajectory design, Ant colony optimization, Planning, Optimization

# 1 INTRODUCTION

The complete automatic design of multiple gravity assist trajectories (MGA), that is the definition of an optimal sequence of planetary encounters and the definition of one or more locally optimal trajectories for each sequence, has been approached with several different techniques. All of them can be classified in two main categories: two level approaches and integrated approaches.

Two-level approaches split the problem into two sub-problems which lay at two different levels: one sub-problem is to find the optimal sequence of planetary encounters; the other is to find an optimal trajectory for that sequence. Two-level approaches define the planetary sequence independently of the trajectory itself. Once the sequence (or a set of promising sequences) has been selected, then the optimal trajectory can be searched for within the set of selected sequences only [1]. Simplified, low fidelity, models for representing the trajectory [2] are used at the first level: this allows for a quick assessment of many sequences, if not all. At the second level, a full model is used to optimize the trajectory. Each sequence is represented by a string of integer numbers, while the associated trajectory is represented with a string of real and integer numbers defining the time and the characteristics of the events occurring along the trajectory (e.g. launch, deep space manoeuvre, arrival at a celestial body, number of revolutions around the Sun, etc.). Therefore, for each sequence, there is an infinite variety of possible trajectories.

The issue with two-level approaches is that it is difficult to assess the optimality of a given planetary sequence without an exhaustive search for all possible trajectories associated with that sequence. Unfortunately, finding an optimal trajectory is a very difficult global optimisation problem in itself. This, combined with the fact that usually there exist a very high number of sequences for a given transfer problem, requires a considerable computational effort. The computational cost can be reduced by discarding non-promising sequences. However, if the low-fidelity model is not accurate enough, either some good sequences are discarded, or many of the retained ones can result to be actually not interesting.

As opposed to the two-level approaches, integrated approaches define a mixed integer-continuous optimization problem, which tackles both the search of the sequence and the optimization of the trajectory by using a single model, at the same time [3]. This kind of problem is known in literature as a hybrid optimization problem [4]. The main difficulty with integrated approaches is that a variation of even a single celestial body in the sequence corresponds to a substantially different set of trajectories. In addition, a variation of the length of the sequence implies varying the number of legs of the trajectory, and thus the total length of the solution vector.

The automatic design of a trajectory with discrete events was recently formulated as a hybrid optimal control problem [5], and a solution was proposed by Conway et al. [6] with a two level approach based on genetic algorithms.

Here we propose to formulate the complete automated design of a multiple gravity assist trajectory as an autonomous planning and scheduling problem. The resulting scheduled plan will provide the planetary sequence for a multiple gravity assist trajectory and a good estimation of the optimality of the associated trajectories.

Although the proposed method can fall in the category of the integrated approaches, the scheduling and the planning of the events are separated at two different levels. A specific MGA trajectory model was developed to automatically schedule the events, if a plan is available, and to provide a good estimation of the feasibility and quality of a trajectory. A novel algorithm, partially inspired by the Ant Colony Optimization (ACO) paradigm [7], was devised to explore the space of possible plans. ACO was originally created to solve the Travelling Salesman Problem [8], and later successfully applied to a number of other discrete optimisation problems. Here the original idea behind ACO was elaborated to solve the planning problem associated to the design of MGA trajectories. The basic ACO paradigm is hybridized with a taboo strategy based on an external archive on infeasible solutions. In the literature, some ACO-derived meta-heuristics exist for the specific solution of different scheduling problems. In particular, Merkle et al. [9] proposed to apply ACO to the solution of the Resource-Constrained Project Scheduling Problem, while Blum, in his work [10], suggested the hybridization of Ant Colony Optimization with a probabilistic version of Beam Search for the solution of the Open Shop Scheduling problem. In this paper, at first we will present the trajectory model and the integrated scheduling of the events, then the novel ACO-based algorithm and how the plan is constructed. Finally, two case studies will demonstrate the effectiveness of the proposed approach at solving known space trajectory design problems.

## 2    TRAJECTORY MODEL

The trajectory model is based on a two-dimensional linked-conic approximation of the trajectory and of the orbits of the planets. The trajectory is composed of a sequence of planar conic arcs linked together through discrete, instantaneous events. In particular, the sequence is continuous in position and piecewise continuous in velocity, i.e. each event introduces a discontinuity in the velocity of the spacecraft but not in its position. The discrete events can be: launch, deep space manoeuvre (DSM), swing-by, and brake.

A final assumption of the present implementation is that all the orbits of both spacecraft and celestial bodies are direct, thus no retrograde orbits are allowed.

In summary, the proposed trajectory model is composed of: a launch from the departure celestial body; a series of deep space flight legs connected through gravity assist manoeuvres (modelled through a linked-conic approximation); an arrival at a target celestial body. Each one of these basic components will be explained in the following sections.

### 2.1    Launch

The launch event is modelled as an instantaneous change of the velocity of the spacecraft with respect to the departure planet. The velocity change is given in terms of modulus $v_0$ (which depends on the capabilities of the launcher) and in-plane direction, specified through the angle $\varphi_0$, measured counter clockwise with respect to the planet's orbital velocity vector $\mathbf{v}_P$ at time of launch $t_0$.

$t_0$ and $\varphi_0$ are free parameters of the model, while launch velocity modulus $v_0$ will be used to target the next planetary encounter and solve the phasing problem, as explained later.

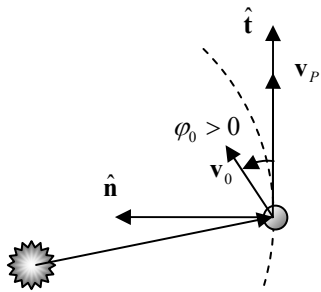

**Fig. 1: Geometry of the launch, and convention for launch angle.**

## 2.2 Swing-by

Gravity assist manoeuvres, or swing-bys, are modelled as instantaneous changes of the velocity vector of the spacecraft due solely to the gravity field of the planet.

Given the relative velocity vector $\mathbf{v}_\infty^-$ before the swing-by, the physical properties of unperturbed hyperbolic orbital motion prescribe that $v_\infty^+ = v_\infty^- = v_\infty$, which means that the modulus of the outgoing velocity $v_\infty^+$ at infinity is known. Its direction can be computed considering the anomaly of the outgoing asymptote (see Fig. 2):

$$\theta_\infty = \arccos\left(\frac{-\mu_P/r_p}{v_\infty^2 + \mu_P/r_p}\right). \tag{1}$$

Here, $\mu_P$ is the gravity constant of the planet, and $r_p$ is the radius of the pericentre of the hyperbola. The value of $r_p$ can be used to control the magnitude of the deflection angle $\delta = \pm(2\theta_\infty - \pi)$ of the incoming velocity and is limited to be above the radius of the planet, $R_P$, to avoid a collision, or to be above the atmosphere to avoid a re-entry. The direction of deflection is determined using a signed radius of pericentre $r_{ps}$, such that $r_p = |r_{ps}|$ and $\delta = \mathrm{sgn}(r_{ps})(2\theta_\infty - \pi)$.

Once $\delta$ is computed, the relative outgoing velocity is calculated by rotating $\mathbf{v}_\infty^-$ in the plane of an angle $\delta$. As for the launch velocity magnitude, the radius of pericentre $r_{ps}$ is tuned to meet the terminal conditions of the transfer leg following the swing-by.

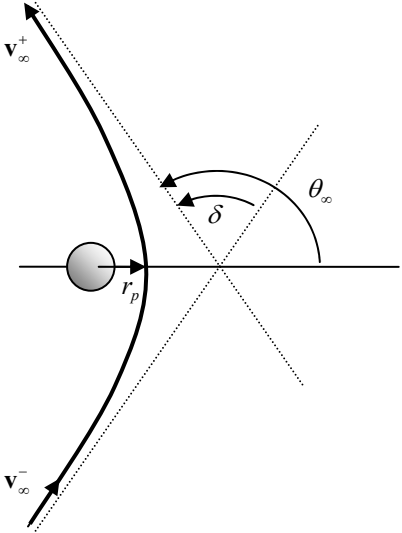

**Fig. 2: Geometry of the swing-by.**

## 2.3 Deep space flight leg

Each deep space flight leg starts at a departure planet $P_i$ and ends at an arrival planet $P_{i+1}$, and is made of two conic arcs linked at a point $M_i$. If the leg contains a deep space manoeuvre, this is applied in this point, and it produces an instantaneous change in the heliocentric velocity vector of the spacecraft, due to an ignition of the engines. In this model, we assume that the DSM is performed either at the apocentre or pericentre of the conic arc preceding the manoeuvre. In addition, the change in velocity is tangential to that arc.

For clarity, in the remainder of this section, we neglect the subscript index *i* of the leg in all the variables.

### 2.3.1 First arc

Let us assume that the spacecraft is at a given planet $P_1$ at time $t_1$. Its position coincides with that of the planet, which is known from the 2D ephemeris. The heliocentric velocity of the spacecraft, instead, depends on the preceding launch or swing-by event.

If the transfer leg contains a DSM, the first step is to find the position *M* and time $t_{DSM}$ of the deep space manoeuvre. The position can either be the pericentre or the apocentre, according to a binary parameter $f_{p/a}$. The true anomaly of the DSM is given by

$$f_{p/a} = \begin{cases} 0 & \to \quad \theta_{DSM} = 0 \\ 1 & \to \quad \theta_{DSM} = \pi \end{cases}.$$

The time of the DSM $t_{DSM}$ is found by using the Kepler's time law [11]. The parameter $n_{rev,1}$ is used to count the number of full revolutions before the deep space manoeuvre.

At *M*, the DSM is applied tangentially, being the free parameter $m_{DSM}$ the magnitude and direction of the DSM: if $m_{DSM}$ is positive, the thrust is along the velocity of the spacecraft, otherwise it is against the velocity of the spacecraft. The complete state of the spacecraft at the beginning of the second arc is thus fully determined.

Note that the use of a tangential DSM at apsidal points is one of the simplifying assumptions of the trajectory model, which allows reducing the number of free parameters. This type of DSM is particularly effective for perigee/apogee raising/decreasing, or period change.

If the leg does not contain any DSM, i.e. $m_{DSM} = 0$, the first arc is propagated up to a fictitious point *M*, defined by adding an angle $\Delta_\theta$ to the initial true anomaly of the spacecraft. The reason for using this forced propagation is twofold: first, we want to force the spacecraft to move away from the planet, before computing the second leg. This is done to prevent that, if no full revolutions are considered, the first intersection could happen after a null time. Second, it prevents any event (swing-by or DSM) to happen immediately after the swing-by, which would be unfeasible due to spacecraft operation constraints. Therefore, the quantity $\Delta_\theta$ has to be larger than the machine numerical precision but small enough to allow for the modelling of short transfer legs. It is important to underline the value of $\Delta_\theta$ is not a design parameter of the model and does not affect the efficiency of the optimization process. For this work, a value of $\Delta_\theta = 0.3$ rad (about 17 deg) was chosen. For this work, $\Delta_\theta = 0.3$ rad was chosen. The time $t_M$ at *M* is found by solving again the Kepler's time law. In this case, parameters $f_{p/a}$ and $n_{rev,1}$ are not used.

### 2.3.2 Second arc

The second arc starts at point *M* and is propagated until the intersection of the orbit of planet $P_2$. Given the orbital parameters of the spacecraft at *M*, and the orbital parameters of planet $P_2$, the task is to find the intersection between the two coplanar orbits. If there are no intersections, the leg is unfeasible, and the initial conditions of the leg, or its parameters, have to be changed. Otherwise, one of the two possible intersections is selected according to the binary parameter $f_{1/2}$: let us call $\theta_{int}, \bar{\theta}$ the true anomalies of the selected intersection, respectively along the orbit of the spacecraft and of the planet. From $\theta_{int}$, the time $t_{int}$ at which the spacecraft intersects $P_2$'s orbit can be computed with the time law, and considering the integer parameter $n_{rev,2}$ counting the number of full revolutions between the point $M$ and the orbital intersection. Figure 3 illustrates a complete leg, including a DSM. The figure highlights that the orbital intersection does not imply, in general, that the planet is at the intersection point at the correct time. This issue will be addressed in the following paragraph.

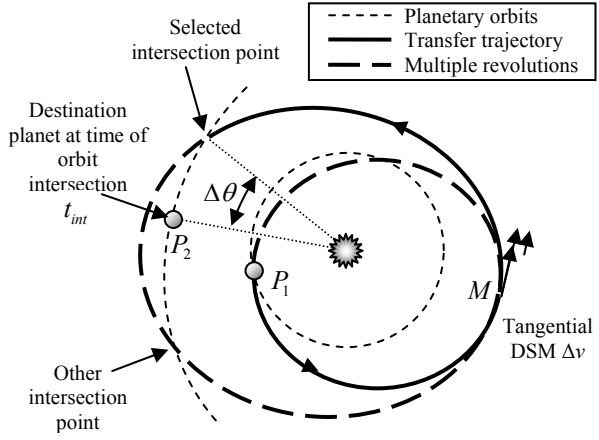

Fig. 3: Representation of a complete leg, starting from $P_1$ with either a swing-by or launch, with a DSM and possibly multiple revolutions. The phasing problem at $P_2$ is not solved, as $P_2$ at the time of intersection is not at the intersection point.

## 2.4 Solution of the Phasing Problem

In order to perform a gravity assist manoeuvre or a planetary capture, the terminal position of the spacecraft has to match that of the planet. However, at intersection time $t_{int}$, planet $P_2$ is at true anomaly $\theta_{P_2}$, which is generally different from $\bar{\theta}$. The time of intersection is a function of the states at the beginning of the leg, which ultimately depend on $v_0$ or $r_{ps}$ depending on the starting event. Therefore, if we introduce the symbol $\lambda$, such that $\lambda \equiv r_{ps}$ if swing-by, or $\lambda \equiv v_0$ if launch.

The true anomalies of the intersection point and of the planet can be expressed as $\bar{\theta}(\lambda)$ and $\theta_{P_2}(\lambda)$. Matching the position of the planet with that of the intersection point at time $t_{int}$ (also known as the phasing problem), then, translates into finding a value $\lambda = \lambda^*$ that satisfies the equation (see Fig. 4):

$$\Delta\theta(\lambda^*) \equiv \theta_{P_2}(\lambda^*) - \bar{\theta}(\lambda^*) = 0 \tag{2}$$

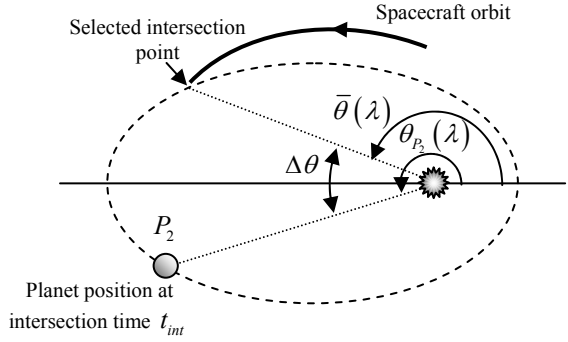

**Fig. 4: The phasing problem consists of finding $\lambda$ such that the target planet $P_2$ is at the orbital intersection point at the correct time. This is done by finding the zero of the difference in true anomalies $\Delta\theta$.**

Fig. 5 (a) and (b) represent the function $\Delta\theta(\lambda)$ for different transfer cases. The non-resonant case depicted in Fig. 5 (a) shows that the function $\Delta\theta(\lambda)$ is continuous, smooth and monotonic over the range of interest of $\lambda$. Hence, the phasing problem has only one solution. This solution can be found with a simple Newton-Raphson method in one dimension. However, when a resonant transfer is considered, as in Fig. 5 (b), $\Delta\theta(\lambda)$ is discontinuous and multiple zeros exist. Each zero corresponds to a different resonance with the planet (and of course a different transfer time). Since there is no easy way, at a given transfer, to prefer one value of $\lambda^*$ over another, all the solutions need to be retained for the evaluation of the following transfers.

In ACO-MGA, the search for the zeros of the function $\Delta\theta(\lambda)$ is performed with the Brent method. A set of starting points, defining multiple intervals for the bisection method, needs to be provided to initialize the Brent method and are specified case by case.

Note that in the examples in Fig. 5, the parameter $\lambda$ is the launch excess velocity $v_0$. It is possible to show that the same behaviour of $\Delta\theta(\lambda)$ holds for a leg starting with a swing-by (i.e. $\lambda \equiv r_{ps}$).

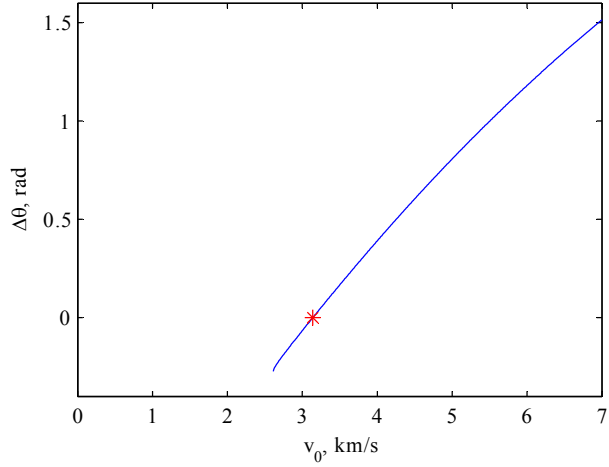

a)

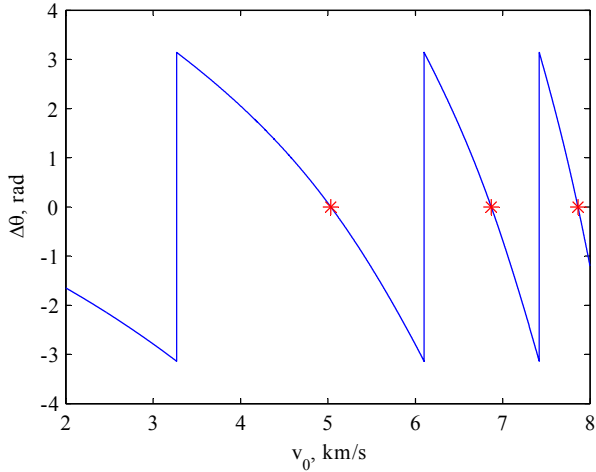

b)

**Fig. 5: Δθ(v₀) for: (a) Earth to Venus leg following launch from Earth. (b) Earth to Earth leg following launch from Earth.**

## 2.5   Complete trajectory

A complete trajectory is made of a sequence of transfer legs connecting $n_P+1$ celestial bodies $\left[P_0, P_1, \ldots P_{n_P}\right]$. The trajectory starts from $P_0$ at time $t_0$ with a launch event characterized by a departure angle $\varphi_0$. The first leg goes from $P_0$ to $P_1$, then a number of swing-bys and interplanetary legs follow, until the final planet $P_{n_P}$. Thus, a complete trajectory with $n_P+1$ planets has $n_P$ legs, and $n_P-1$ swing-bys.

The solution of Eq. (2) provides a complete scheduling of the trajectory given the initial time $t_0$ and the five parameters $\left[m_{DSM}, n_{rev,1}, n_{rev,2}, f_{p/a}, f_{1/2}\right]_i$ for every leg $i = 0, \ldots, n_P - 1$.

Since these five parameters fully characterize all possible legs from a planet $P_i$ to a planet $P_{i+1}$, they are said to define a *type of transfer*. Conversely, because of the multiplicity of the zeros of Eq. (2), each type of transfer corresponds to a set of trajectories.

Hence, assigning a value to $t_0$, $\varphi_0$, $P_i$, $P_{n_P}$, $\left[m_{DSM}, n_{rev,1}, n_{rev,2}, f_{p/a}, f_{1/2}\right]_i$ for $i = 0,...,n_P - 1$ creates a plan, or *solution*, which is a tree structure in which every branch, from root to leaves, is a *trajectory*. Each trajectory is characterised by a different combination of $v_0^*$ and $r_p^*$ for each leg.

The entire tree is a complete set of trajectories from $P_0$ to $P_{n_P}$ and represents a solution of the MGA trajectory planning problem. Thus, a plan is fully defined by assigning a value to the parameters in Table 1 for all legs $i = 0,...,n_P - 1$. Algorithm 1 keeps track of all the trajectories in the tree, and yields a list $L$ containing all the possible conditions of arrival at the last reachable planet. If no trajectory in the set associated to leg $i$ satisfies the phasing problem, then planet $i+1$ cannot be reached and the algorithm terminates. A partial or incomplete solution is the set of parameters sufficient to describe a solution up to leg $i$. Furthermore, if no solution to the phasing problem exists at leg $i$, the plan is broken and the solution is said to be infeasible at leg $i$. Furthermore, an upper bound on the time of flight of the entire trajectory, or of some legs, is introduced. Trajectories that exceed the total or partial time of flight constraint are discarded from the list. The information of infeasibility at a given transfer will be used to fill in a taboo list of broken or incomplete solutions.

For each of the trajectories found, the model computes: the sum of all the deep space manoeuvres, or total $\Delta v$ and the launch excess velocity, $v_0$; the relative velocity at the last planet, $v_\infty$; the total time of flight of the trajectory, $T$. The objective value $f_{obj}$ of the trajectory depends on the problem and it is usually a function of these values.

The whole model was implemented in ANSI C and compiled as a MEX-file for interfacing with MATLAB.

**Table 1: Description of the free design variables defining a solution according to the proposed 2D model.**

| Description | Variables |
| --- | --- |
| Planetary sequence | $\left[P_0, P_1, ... P_{n_P}\right]$ |
| Departure time | $t_0$ |
| Departure angle | $\varphi_0$ |
| Types of transfer for $i = 0,...,n_P - 1$ | $\left[m_{DSM}, n_{rev,1}, n_{rev,2}, f_{p/a}, f_{1/2}\right]_i$ |

**Algorithm 1: Generation of a list *L* containing the arrival conditions for all the feasible trajectories of the transfer problem.**

| | |
|---|---|
| 1: | On the first leg, find all possible $v_0^* \mid \Delta\theta(v_0^*) = 0$ |
| 2: | **For** each $v_0^*$ find the final conditions of the leg at planet $P_1$ |
| 3: |     Add all the possible final conditions to list *L* |
| 4: | **End For** |
| 5: | **For** each leg $i = 0,...,n_P - 1$ |
| 6: |     $L_{temp} \leftarrow \emptyset$ |
| 7: |     **For** each element in list *L* |
| 8: |         Find all possible $r_{ps}^* \mid \Delta\theta(r_{ps}^*) = 0$ |
| 9: |         **For** each $r_{ps}^*$ |
| 10: |             Find the final conditions at planet $P_{i+1}$ and add to list $L_{temp}$ |
| 11: |         **End For** |
| 12: |     **End For** |
| 13: |     **If** $L_{temp} = \emptyset$, **Then** All trajectories infeasible at leg *i*, **Terminate** |
| 14: |     $L \leftarrow L_{temp}$ |
| 15: | **End For** |

## 3   THE ACO-MGA ALGORITHM

The model described in the previous section yields a set of scheduled trajectories provided that a complete or partial plan is available. In this section, we present an optimization procedure, based on the ant colony optimization paradigm, to explore the space of possible plans.

At first, the continuous space of the real parameters $t_0$, $\varphi_0$ and $m_{DSM}\vert_i$ is reduced to a finite set of states. Then, the optimization algorithm, called ACO-MGA in the following, operates a search in the finite space of possible values for the design parameters. A complete description of the algorithm ACO-MGA follows.

### 3.1   Solution coding

In ACO-MGA, a solution is coded through a string of discrete values assigned to the parameters. However, the set of parameters discussed before is inhomogeneous, as it is made of real, integer and binary variables. In particular, $t_0$, $\varphi_0$ and $m_{DSM}\vert_i$ are real continuous variables and need to be properly discretised. In the present implementation of ACO-MGA, the values of the departure date $t_0$ and the departure angle $\varphi_0$, are assumed to be pre-assigned and therefore the two parameters are removed from the list of the variables. The rationale behind this choice is that, although the launch date has a great impact on the resulting trajectory, if an algorithm exists that is able to efficiently generate a complete plan for a given launch date, a systematic search can be performed along the launch window, with a given time step (as will be shown in one of the test cases). The angle $\varphi_0$, on the other hand, is often dictated by the launch capabilities and the first planetary encounter [1]. $m_{DSM}$ is a design parameter in the optimisation: the discretisation is chosen *a priori* and it can be, for example, uniform within a range. The maximum value (in magnitude) depends on the engine capabilities and on the transfer problem itself. It is a task of the user to choose a significant set of values for this parameter.

Using the additional assumptions on $t_0$, $\varphi_0$, and fixing $P_0$, each solution can be coded using a vector $\mathbf{s}$ of positive integers. The vector has $2n_{legs}$ components. Each pair of consecutive components encodes all the parameters necessary to characterise one leg of the solution (Fig. 6). The first element of the pair is encoding the identification number of the target planet for the corresponding leg according to the following procedure: an ordered list $\mathbf{q}_{P,i}$ containing all the planets available as a target for each leg $i$ is predefined (and fixed); then, let $k = s_{2(i-1)+1}$, the target planet is the $k^{th}$ entry in the list $\mathbf{q}_{P,i}$, i.e. $q_{P,(i,k)}$.

The second element of the pair is the row index of a matrix $\mathbf{G}_i$ containing all possible combinations of indexes identifying the elements of the five sets: $\mathbf{q}_{1,i}, \mathbf{q}_{2,i}, \mathbf{q}_{3,i}, \mathbf{q}_{4,i}, \mathbf{q}_{5,i}$. These sets contain the possible values for each one of the five parameters identifying the type of transfer at leg $i$. Thus, each row of $\mathbf{G}_i$ is a vector representing a different type of transfer. The matrix $\mathbf{G}_i$ has $|\mathbf{q}_{1,i}| \times |\mathbf{q}_{2,i}| \times |\mathbf{q}_{3,i}| \times |\mathbf{q}_{4,i}| \times |\mathbf{q}_{5,i}|$ rows and 5 columns ($|\cdot|$ is the cardinality of a set). The parameters for the $j^{th}$ type of transfer for the $i^{th}$ leg can be obtained as follows:

$$m_{DSM}|_i = q_{1,(i,k_1)}$$
$$n_{rev,1}|_i = q_{2,(i,k_2)}$$
$$n_{rev,2}|_i = q_{3,(i,k_3)}$$
$$f_{p/a}|_i = q_{4,(i,k_4)}$$
$$f_{1/2}|_i = q_{5,(i,k_5)}$$

where $k_l = G_{i,(j,l)}$, $l = 1,...,5$. If $\mathbf{q}_{4,i} = \mathbf{q}_{5,i} = \{0,1\}$, then the matrix $\mathbf{G}_i$ is:

$$\mathbf{G}_i = \begin{bmatrix} 1 & 1 & 1 & 1 & 1 \\ 1 & 1 & 1 & 1 & 2 \\ 1 & 1 & 1 & 2 & 1 \\ 1 & 1 & 1 & 2 & 2 \\ 1 & 1 & 2 & 2 & 1 \\ 1 & 1 & 2 & 2 & 2 \\ \vdots & \vdots & \vdots & \vdots & \vdots \\ |\mathbf{q}_{1,i}| & |\mathbf{q}_{2,i}| & |\mathbf{q}_{3,i}| & 2 & 2 \end{bmatrix}$$

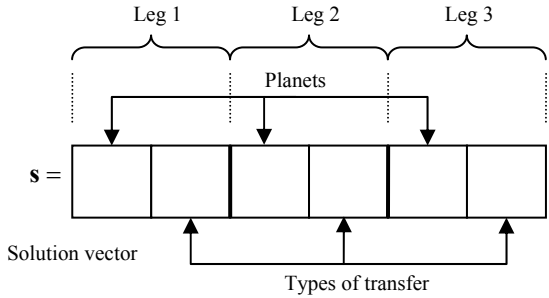

Fig. 6: Solution vector s for coding a three-leg solution.

### 3.2 The taboo and feasible lists

A transfer from planet $P_i$ to planet $P_{i+1}$ can be feasible or unfeasible, for the same set of parameters, depending on all the preceding legs from 0 to $(i-1)$. For this reason, when an infeasible leg is generated, it is necessary to store the path that led to that infeasible leg. Thus, all the parameters characterising the partial solution up to $P_{i+1}$ are stored in a taboo list.

In particular, if the problem involves $n_P$ legs, then the same number of taboo lists are used. The taboo list $L_{taboo,i}$ of leg $i$ contains all the partial solutions which are unfeasible at leg $i$ (but feasible for legs $1...i-1$). Each taboo list consists of a matrix, which has an arbitrary number of rows and $2i$ columns.

Dual to the list of taboo partial solutions, the feasible list $L_{feas}$ stores all the solutions, which are completely feasible, i.e. reach $P_{n_P}$. This is once more a matrix with an arbitrary number of rows and $2n_P$ columns.

Since each solution contained in the feasible list is complete, then it is possible to associate an objective value (or cost) to each one of them. In the following test cases, we will use, as objective value, the $v_\infty$ and a combination of $v_\infty$ and $T$. Note that, since, in general, there is more than one trajectory for a given solution (i.e. for a given set of free design variables), the objective value of a solution is given by the best trajectory value.

### 3.3 Search engine

The search space is organised as an acyclic oriented tree. Each branch of the tree represents a leg of the problem, while each node (or leave) represents a different destination planet and type of transfer. A population of virtual ants are dispatched to explore the tree, searching for an optimal solution.

The search runs for a given number of iterations $n_{iter,max}$, or until a maximum number of objective function evaluations $n_{eval,max}$ has been reached. An evaluation is a call to the model, in order to compute the objective value associated to a given solution.

Algorithm 2 illustrates the main iteration loop. Each iteration consists of two steps: first, a solution generation step, and then a solution evaluation step. In the former step, the ants incrementally compose a set of solution vectors, while the latter invokes the trajectory model to assess the feasibility and the objective value of each generated solution. When the main loop of the search stops, the feasible list $L_{feas}$ contains all the solutions, which were found feasible, with their corresponding objective value. The solutions are then sorted according to their objective value.

**Algorithm 2: Main ACO-MGA search engine.**

```
1  : While n_iter < n_iter,max ∧ n_eval < n_eval,max, Do
2  :     For each ant k = 1…m
3  :         s ← Generate planetary sequence
4  :         s ← Generate types of transfers
5  :         If s is not discarded, S ← S ∪ {s}
6  :     End For
7  :     Evaluate all solutions in S
8  :     Update L_feas and L_taboo,i
9  :     Update n_iter, n_eval
10 : End Do
11 : Sort L_feas according to f_obj.
```

## 3.4 Solution generation

The tree is simultaneously explored, from root to leaves, by *m* ants. At each iteration, each one of the *m* ants explores the tree independently of the others, but taking into account the information collected by all the ants, at previous iterations, and stored in the feasible and taboo lists. As an ant moves along a branch, it progressively composes a complete solution. At first, each ant assigns a value to the odd entries of the solution vector, i.e. composes the sequence of planetary encounters, then it assigns a value to the even entries of the solution vector, i.e. the parameters defining the type of transfer for each legs.

### 3.4.1 Planetary sequence generation

The process is described in pseudo-code in Algorithm 3. Each ant composes a solution adding one planet at the time. As the departure planet is given, the ant has only to choose the destination planet for each leg. The choice is made probabilistically by picking from the list $\mathbf{q}_{P,i}$. The selection depends on the discrete pheromone distribution vector $\boldsymbol{\tau}_{P,i}$ (one for every leg) which contains the pheromone level associated to each body in $\mathbf{q}_{P,i}$. Note that we use the same notion of pheromone as in standard ACO [7], however there are some differences. Here, the pheromone level of each possible choice at each leg depends on the previous legs, and therefore it is computed at every step. Furthermore, due to the different pheromone update rule, here the amount of pheromone is not upper limited to 1. Every time an ant is at leg *i*, the pheromone distribution vector is reset to $\boldsymbol{\tau}_{P,i} = \begin{bmatrix} 1 & 1 & \ldots & 1 \end{bmatrix}^T$. As it will be explained, this is equivalent to state that all the planets have equal probability to be chosen. The ant sweeps the entire list $\mathbf{q}_{P,i}$ substituting the identification number of each element in $\mathbf{q}_{P,i}$ into the $i^{th}$ odd component of the partial solution vector $\mathbf{s}$. Then, the feasible list is searched, for all the solutions which have a (partial) planetary sequence which matches the one in $\mathbf{s}$. Say that the $j^{th}$ element of $\mathbf{q}_{P,i}$ is added to $\mathbf{s}$, and the resulting partial sequence in $\mathbf{s}$ matches the partial sequence of the $l^{th}$ solution in the feasible lists, then the pheromone level $\tau_{P,(i,j)}$ associated to the $j^{th}$ element of $\mathbf{q}_{P,i}$ is increased as follows:

$$\tau_{P,(i,j)} \leftarrow \tau_{P,(i,j)} + \frac{1}{f_{obj,l}} w_{planet} \tag{3}$$

The amount of pheromone which is deposited depends on the objective value $f_{obj,l}$ of the matching solution in the feasible list, and on the weight $w_{planet}$. Once the pheromone update has been done for all the possible choices, the probability of selecting one of them is given by $p_{P,(i,j)} = \tau_{P,(i,j)} / \sum_j \tau_{P,(i,j)}$, and a random selection is performed according to this probability distribution. Thus, the probability of choosing the $j^{th}$ planet increases according to how many times it generates a promising sequence (leading to a feasible solution), to the value of the feasible solution itself, and to the parameter $w_{planet}$. This mechanism is analogous to the pheromone deposition of standard ACO and aims at driving the search of the ants toward planetary sequences, which previously led to good solutions. In fact, those planets which generate (partial) sequences that appear either frequently in the feasible list, or rarely, but with low objective function are selected with higher probability. On the other hand, the probability of selecting other planets remains positive, such that one or more ants can probabilistically choose a planet that generates an undiscovered sequence. Note that, if the feasible list is empty, then all the planets have the same probability to be selected.

The parameter $w_{planet}$ controls the learning rate of the ants. A low value of $w_{planet}$ will make the term $w_{planet}/f_{obj,l}$ small, and thus the probability distribution will not change much, even if the solution appears repeatedly in the feasible list, or with low values of $f_{obj}$. Thus, a relatively low value of $w_{planet}$ will favour a global exploration of the search space, while a high value of $w_{planet}$ will greatly increase the probability of choosing a planet which led to a feasible sequence. If the value of $w_{planet}$ is high enough with respect to a reference objective value, then the ant will preferably choose a feasible sequence, rather than trying a new one, which has not proven to be feasible. For these reasons, we can say that low values of $w_{planet}$ will favour local exploration of planetary sequences.

The procedure iterates for all the legs of the problem, and for all the ants. At the end, all the odd entries of the temporary solution **s** contain a target planet and the planetary sequence is complete. The next step is to find the type of transfers for each leg, thus filling the even entries of **s** and complete the solution.

**Algorithm 3: Generation of the planetary sequence of the temporary solution probabilistically.**

| | |
|---|---|
| 1: | **For** each leg $i = 1, \ldots, n_P$ |
| 2: | $\tau_{P,i} \leftarrow [1 \; 1 \; \ldots \; 1]$ |
| 3: | **For** each target body $j$ available at leg $i$ |
| 4: | $s_{temp}(1+2(i-1)) \leftarrow j$ |
| 5: | **For** each solution $l$ in $L_{feas}$ that matches $\mathbf{s}_{temp}$ |
| 6: | $\tau_{P,(i,j)} \leftarrow \tau_{P,(i,j)} + \dfrac{1}{f_{obj,l}} w_{planet}$ |
| 7: | **End For** |
| 8: | **End For** |
| 9: | $s_{temp}(1+2(i-1)) \leftarrow \text{RandomSelection}(\tau_{P,i})$ |
| 10: | **End For** |

### 3.4.2 Type of transfer generation

Once an ant has filled in the odd components of a solution **s**, it proceeds assigning values to the even components (see Algorithm 4). Similarly to the planet sequence generation, for each transfer all the available types of transfer are assigned, one at the time, to the solution **s**. A vector **s** for which a value is assigned to both the odd and even components up to leg $i$ represents a partial solution. Similarly to before, a vector $\boldsymbol{\tau}_{t,i}$ contains the pheromone associated to the rows of the matrix $\mathbf{G}_i$ (i.e. to each type of transfer).

For each new partial solution, the taboo list is first checked. If the partial solution appears in the taboo list, then it means that this solution will be infeasible, regardless of the way it is completed. The pheromone of the type of transfer associated to that sequence is set to zero, to avoid future selection of that type of transfer. If the partial solution does not appear in the taboo list, the feasible list is searched for any matching partial solution. For every match found, the pheromone of that type of transfer is modified as follows:

$$\tau_{t,(i,j)} \leftarrow \tau_{t,(i,j)} + \frac{1}{f_{obj,l}} w_{type} \tag{4}$$

where the weight $w_{type}$ is introduced with analogous meaning to $w_{planet}$. In fact, the higher the coefficient, the higher the chances that solutions similar to the feasible ones are generated. Conversely, a low value of $w_{type}$ will favour the selection of sequences with a different type of transfer, thus increasing the random exploration of the whole solution space. The selection of the type of transfer happens probabilistically, in the same way as for the planet before.

If, at a given leg $i$, all possible transfer types correspond to partial solutions in the taboo list, the vector of pheromone distribution $\boldsymbol{\tau}_{t,i}$ will be full of zeros. As a consequence, the solution **s** (which can be partial or complete) is discarded, and the ant can stop its exploration of that branch of the tree. At the end of the solution generation step, the solution **s** is either discarded or completed. Once all the ants have completed their exploration, the result is a number of solutions (less than or equal to the number of ants $m$) to be evaluated.

**Algorithm 4: Generation of the types of transfer of the temporary solution probabilistically.**

| | |
|---|---|
| 1: | **For** each leg $i = 0, \ldots, n_P - 1$ |
| 2: | $\boldsymbol{\tau}_{t,i} \leftarrow \begin{bmatrix} 1 & 1 & \ldots & 1 \end{bmatrix}$ |
| 3: | **For** each type of transfer $j$ available for leg $i$ |
| 4: | $s_{temp}(2 + 2(i-1)) \leftarrow j$ |
| 5: | **If** $\mathbf{s}_{temp} \in L_{taboo,i}$ taboo list of leg $i$ **Then** |
| 6: | $\tau_{t,i}(j) \leftarrow 0$ |
| 7: | **Else** |
| 8: | **For** each solution $l$ in $L_{feas}$ that matches with $\mathbf{s}_{temp}$ |
| 9: | $\tau_{t,i}(j) \leftarrow \tau_{t,i}(j) + \frac{1}{f_{obj,l}} w_{type}$ |
| 10: | **End For** |
| 11: | **End If** |
| 12: | **End For** |
| 13: | **If** $\sum_j \tau_{t,i}(j) = 0$ **Then** |
| 14: | Discard this solution, **Terminate** |
| 15: | **Else** |
| 16: | $\mathbf{s}_{temp}(2 + 2(i-1)) \leftarrow \text{RandomSelection}(\boldsymbol{\tau}_{t,i})$ |
| 17: | **End If** |
| 18: | **End For** |

### 3.5 Solution evaluation

Once a set of solutions **S** has been generated by the ants, each solution has to be evaluated to assess its feasibility and its objective value. This is done by calling the trajectory model.

Solutions in **S** are evaluated one by one, by means of the model presented before. The trajectory model can be seen as a function which takes a solution vector **s** as an input, together with $t_0, \varphi_0, P_0$, and gives as an output either an objective value $f_{obj}$ (if the solution is feasible) or the leg $l_u$ at which the solution becomes unfeasible. If the solution is feasible, it is stored in the feasible list and $l_u = 0$, otherwise it is stored in the $l_u{}^{th}$ taboo list.

### 3.6 Differences with standard ACO algorithm

Although the proposed approach takes its inspiration from ACO, and substantially applies the same principle, there is a fundamental difference that prevents the use of the standard ACO for the MGA problem. In the MGA problem, as opposed to the TSP tackled with ACO, the pheromone cannot be assigned to single legs of the trajectory: this is due to the fact that each leg (identified by its couple of integers) has no intrinsic value in the trajectory, if disconnected from the previous legs. In other words, the only two parameters (planet and type of transfer) are not sufficient to fully characterise one leg: the actual value of the leg depends also on its initial conditions, which are in turn determined by the parameters of the previous legs. As a comparison, in the TSP, the distance between pair of cities, that is summed up to build the tour cost, is constant and fixed a-priori, for a given instance of the problem. For this reason, it is not possible to assign pheromone to single legs, independently from the rest of the solution.

## 4 CASE STUDIES

The proposed optimisation method was applied to two case studies inspired by the Laplace [12] and Cassini [13] missions, the former of which currently under preliminary study by ESA, NASA and JAXA.

ACO-MGA was tested against genetic algorithms, which are known to perform well on problems with discrete variables. In particular, it was chosen to use the genetic algorithm implemented in MATLAB® within the Genetic Algorithm and Direct Search Toolbox™ (GATBX), and the Non-dominated Sorting Genetic Algorithm (NSGA-II) [14]. Settings for all the optimisers will be specified for each test case. While NSGA-II can deal with discrete variables, GATBX only uses real variables: a wrapper of the objective function was coded to round the continuous solution vector to the closest integer.

Due to the stochastic nature of the methods involved in the comparative tests, all the algorithms were run 100 times. The performance index used to compare the ACO-MGA against the other global optimisers is the success rate for a given number of function evaluations: according to the theory developed in [15], 100 repetitions give an error in the determination for the exact success rate of less than 6%, with 92% confidence.

Some preliminary tests showed that the best performances of ACO-MGA are achieved if the algorithm is run in 2 steps, using different sets of parameters. In particular, in the first step, the weights $w_{planet}, w_{type}$ are set to null: remembering Eq. (3) this choice translates into an initial pure random search. In the second step, weights are set to non-null values, to explore around the feasible solutions found.

The values of $w_{planet}, w_{type}$ are set such that:

$$w_{planet}, w_{type} = \bar{w} \cdot \tilde{f}_{obj} \tag{5}$$

where $\tilde{f}_{obj}$ is the expected value for the objective function. In this way, choosing for example $\bar{w}=1$, a 1 is added to the pheromone of a given element every time a matching solution with objective $\tilde{f}_{obj}$ appears in the feasible list. The value of pheromone is higher if the objective value of the matching feasible solution is lower.

This two-step procedure can be explained in the following way. The first step allows a random sampling of the solution space, with the aim of finding a good number of feasible solutions. This is done to prevent the algorithm to stagnate around the first feasible solution found. The second step intensifies the search around the feasible solutions which were found at step one. Because of Eqs. (3) and (4), feasible solutions with low objective value are likely to be investigated further. In addition, the random component in the process does not forbid to keep exploring the rest of the search space.

The test cases were run on an Intel® Pentium® 4 3 GHz machine running Microsoft® Windows® XP.

### 4.1 Laplace case study

In this mission, the spacecraft reaches the sphere of influence of Jupiter after an interplanetary flight, and exploits a swing-by of Ganymede to get captured into the Jovian system. At this point, multiple swing-bys of Ganymede and Callisto are used to reduce the relative velocity to Callisto $v_\infty$, in order to be captured by the moon and start the scientific phase of the mission.

The problem under consideration relates to the second part of the transfer: it is assumed that the interplanetary trajectory has been already optimised, including the first Ganymede gravity assist. The resulting orbit has a 3:1 resonance (spacecraft:planet) with Ganymede. The problem is to find the sequence of additional swing-bys, starting from the second one of Ganymede, to reach Callisto with low $v_\infty$.

For tackling this problem with ACO-MGA, a launch is simulated from Ganymede, and the initial conditions and type of transfer for the first leg (GG) are kept fixed. The subsequent legs, instead, need to be optimised.

The date of the first Ganymede swing-by is $t_0 = 9309.8$ d, MJD2000, corresponding to 28[th] June 2025, where the spacecraft leaves the planet with an angle $\varphi_0 = 1.2471$ rad. As the first leg is fixed, the associated sets of parameters is:

$$P_1 \in \mathbf{q}_{P,1} = \{\text{Ganymede}\}$$
$$m_{DSM}\big|_1 \in \mathbf{q}_{1,1} = \{0\}$$
$$n_{rev,1}\big|_1 \in \mathbf{q}_{2,1} = \varnothing$$
$$n_{rev,2}\big|_1 \in \mathbf{q}_{3,1} = \{0\}$$
$$f_{p/a}\big|_1 \in \mathbf{q}_{4,1} = \varnothing$$
$$f_{1/2}\big|_1 \in \mathbf{q}_{5,1} = \{1\}$$

Since there is no DSM, the parameters $n_{rev,1}\big|_1 \in \mathbf{q}_{2,1}$ and $f_{p/a}\big|_1 \in \mathbf{q}_{4,1}$ are not relevant. These settings lead to a departure velocity from Ganymede of $v_0 = 5.1$ km/s, as required.

Three free legs are added to the trajectory. For the first two, the algorithm can choose to target Ganymede or Callisto, while for the third and last, the target must be Callisto:

$$P_i \in \mathbf{q}_{P,i} = \{\text{Ganymede, Callisto}\}$$
$$m_{DSM}\big|_i \in \mathbf{q}_{1,i} = \{-10, 0, 10\} \text{ m/s}$$
$$n_{rev,1}\big|_i \in \mathbf{q}_{2,i} = \{0\}$$
$$n_{rev,2}\big|_i \in \mathbf{q}_{3,i} = \{0,1,2,3\} \qquad i = 2,3$$
$$f_{p/a}\big|_i \in \mathbf{q}_{4,i} = \{0,1\}$$
$$f_{1/2}\big|_i \in \mathbf{q}_{5,i} = \{0,1\}$$

$$P_4 \in \mathbf{q}_{P,4} = \{\text{Callisto}\}$$
$$m_{DSM}\big|_4 \in \mathbf{q}_{1,4} = \{-10, 0, 10\} \text{ m/s}$$
$$n_{rev,1}\big|_4 \in \mathbf{q}_{2,4} = \{0\}$$
$$n_{rev,2}\big|_4 \in \mathbf{q}_{3,4} = \{0,1,2,3\}$$
$$f_{p/a}\big|_4 \in \mathbf{q}_{4,4} = \{0,1\}$$
$$f_{1/2}\big|_4 \in \mathbf{q}_{5,4} = \{0,1\}$$

Small corrective DSM manoeuvres of ±10 m/s can be used, and up to 3 complete revolutions can be performed. The number of revolutions is entirely controlled by the parameter $n_{rev,2}\big|_i \in \mathbf{q}_{3,i}$. In general, there is no easy way to identify whether the first or the second orbital intersection is the best one, so the binary parameter $f_{1/2}\big|_i$ was left as a free design variable.

The radius of pericentre of the swing-bys is bounded between 1 and 3 radii $R_P$ of the body. Here a comment is needed: it is usual practice, in preliminary mission design, to consider a safety margin in the closest approach of a planet during a swing-by manoeuvre. Usually this translates in considering a radius of pericentre not smaller than 1.1 $R_P$. The main reason for using a slightly lower minimum altitude is that the number of feasible solutions in the solution space increases, since each swing-by can provide a higher deflection of the velocity vector, and thus the search for the optimal solution results more favourable for all the optimisers, but in particular for the population-based ones. We can also assume that the safety margin can be added when the solution is re-optimised with a more complete model, and assume that a DSM can compensate for the lack of swing-by deflection. Note that this trick will not be adopted in the next, more realistic case study.

The total time of flight was limited to a maximum of 100 days and the objective function for a complete solution is the $v_\infty$ at the final encounter with Callisto.

The average time for evaluating one solution (finding all possible trajectories) is 30.34 ms, and there are 9216 distinct solutions. Thus, a systematic approach, scanning all the solutions, would require 4.66 min.

GATBX and NSGA-II were run for 600 function evaluations with the settings shown in Table 2. The mutation and crossover parameters *pcross_bin* and *pmut_bin* were tuned through several preliminary runs although very little difference in the quality of the results was registered.

Since the size of the population, *popsize*, is very important for genetic-based algorithms, and can significantly affect the results, this case study was also run 100 times with different sizes of the population (maintaining the predefined number of total function evaluations by varying the number of generations accordingly *ngen*). For NSGA-II, it resulted that there was no noticeable change in the quality of the results over 100 runs. This is related to the fact that NSGA-II is not completely converging with only 600 function evaluations. For GATBX, instead, results were changing significantly, and the settings leading to the best solutions were used.

The parameters for ACO-MGA were tuned by running the same test case for different values of the weights, the number of ants and the number of iterations. The maximum number of function evaluations was set to 600 as for the other algorithms although ACO-MGA requires, on average, 545 function evaluations. The best results were obtained with the following settings. 10 ants were used, with a first optimisation step with 30 iterations and $w_{planet}, w_{type} = 0$, followed by a second step of 30 more iterations with $w_{planet}, w_{type} = 20 \cdot 3$ km/s. Because of the normalisation shown in Eq. (5), the non-dimensional weight $\bar{w}$ does not depend on the dimensions and value of the objective function of the specific problem: therefore, we will use the same value also for other transfer problems, as will be shown in the next case study. Note that the two weights $w_{planet}, w_{type}$ are tuning parameters of the optimiser. It was found that the performance of ACO-MGA is not very sensitive to the value of the weight (within reasonable variations). Furthermore, it should be easy, for a given problem, to get a value for $\tilde{f}_{obj}$, either by experience or simply by random sampling the solution space. In the case $\tilde{f}_{obj}$ cannot be determined, then the values of $w_{planet}, w_{type}$ shall be selected by tuning the optimiser on the specific problem.

Results in the form of statistical parameters over the 100 runs are presented in Table 3. All the algorithms found at least one feasible solution in every run. The value of 2 km/s as a target value for the $v_\infty$ has been chosen to compute the success rate according to the procedure proposed in [16].

The results in Table 3 point out that, while all the algorithms find feasible solutions in all the runs, the quality of the solution is much better for ACO-MGA. Moreover, GATBX found a good solution only in 31% of the runs, and NSGA-II in 39%. ACO-MGA, instead, found a good solution in 62% of the runs.

The time for one ACO-MGA run is about 8 min. The simplicity of the test case, together with the implementation of ACO-MGA in a high-level language like MATLAB, makes the use of an optimisation method slower than the systematic scan of the whole solution space. Note that this will not happen in the more complex Cassini test case.

**Table 2: Parameters of GATBX and NSGA-II for the Laplace test case.**

| GATBX | | NSGA-II | |
|---|---|---|---|
| Parameter | Value | Parameter | Value |
| *Generations* | 20 | *ngen* | 22 |
| *PopulationSize* | 30 | *popsize* | 28 |
| *StallGenLimit* | +Inf | *pcross_bin* | 0.5 |
| | | *pmut_bin* | 0.5 |

**Table 3: Comparison of the performances of ACO-MGA, GATBX, NSGA-II over 100 runs for the Laplace problem.**

| | Average best value, km/s | % runs < 2 km/s | % feasible runs |
|---|---|---|---|
| ACO-MGA | 2.0141 | 62% | 100% |
| GATBX | 2.34 | 31% | 100% |
| NSGA-II | 2.1074 | 39% | 100% |

The reference solution for this problem, as chosen by ESA during a preliminary study [17], was re-optimised using a full 3D model with one free deep space manoeuvre per leg [18], and minimising the $\Delta v$: the resulting trajectory is represented in Fig. 7 (a), starting from the second swing-by of Ganymede. The sequence for this solution is GGCGC, and the objective value, i.e. the final relative velocity, is $v_\infty = 1.96$ km/s. The solution is practically ballistic.

ACO-MGA identified a solution, with an objective value of $v_\infty = 1.91$ km/s, equal flyby sequence and comparable transfer times per leg (see Fig. 7 (b)). Table 4 compares some parameters of the 2D solution with the re-optimised 3D solution. Slight differences are due to the different models, and mainly to the changes of inclination that are needed in the 3D solution. When re-optimized with the full 3D model, the 2D solution identified by ACO-MGA yields the reference solution in Fig. 7 (a). This demonstrates that the assumptions underneath the 2D model are consistent and allow for identifying first guess solutions that converge locally to higher fidelity solutions with comparable cost and mission characteristics.

Note that the solution chosen by ESA, and used here as a comparison, is not the best from the point of view of the arrival velocity. In fact, this solution was chosen by ESA following a trade off, taking into account not only the $v_\infty$, but also the presence of DSMs, the total time of flight, the radiation dose, and the arrival velocity vector at Callisto.

ACO-MGA found also solutions with lower $v_\infty$. The best one has $v_\infty = 1.709$ km/s, and corresponds to the trajectory plotted in Fig. 8. The swing-by sequence is the same, GGCGC, but the total time of flight is much longer (92 days), since the first GC leg performs 1 full revolution and the CG leg 2 full revolutions, and a 10 m/s DSM is added to the second leg.

The 100 runs returned a number of feasible solutions. Four different solutions with a low value of $v_\infty$ are described in Table 5, and respectively plotted in Fig. 9. Note that we deliberately decided not to pick the best four solutions according to lowest $v_\infty$, but rather to show the variety of different good solutions that were found through this approach, and demonstrate that this tool could be useful for finding valid alternative options. In the same way, Table 6 and Fig. 10 describe four different solutions among those with short total time of flight. Note that all the solutions with low $v_\infty$ follow the same planetary sequence GGCGC, while the short solutions use also the sequence GGGCC, but at the cost of a higher $v_\infty$.

Finally, the best solution associated to each feasible sequence obtained by ACO-MGA is represented in the bar plot in Table 7.

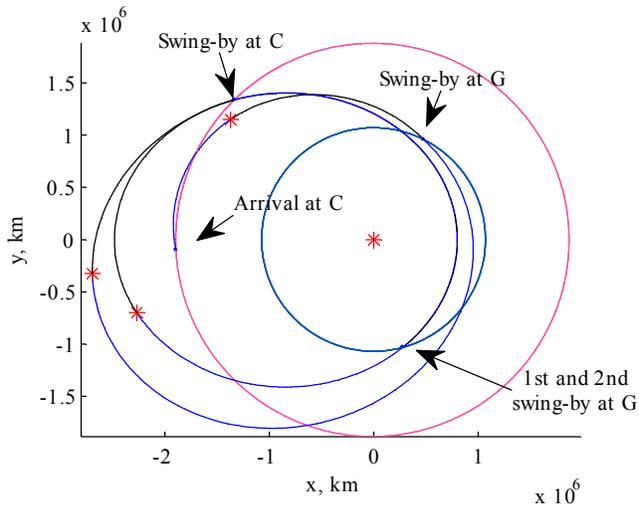

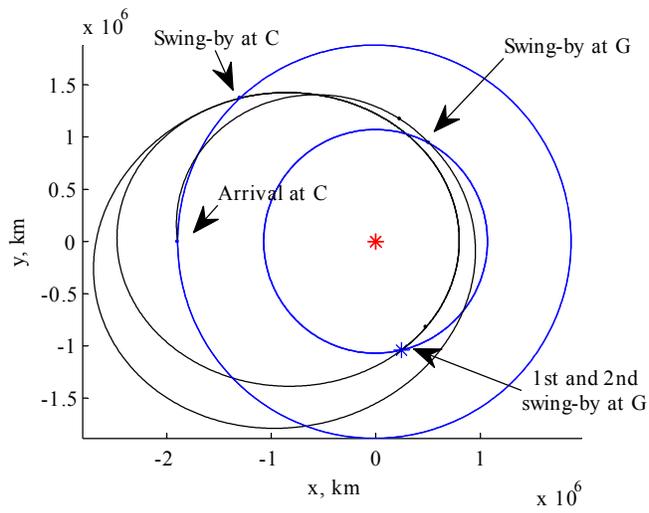

**Fig. 7: (a) Reference solution (sequence GGCGC) optimised with a full 3D model. (b) Same solution as found by ACO-MGA. The first leg is not plotted.**

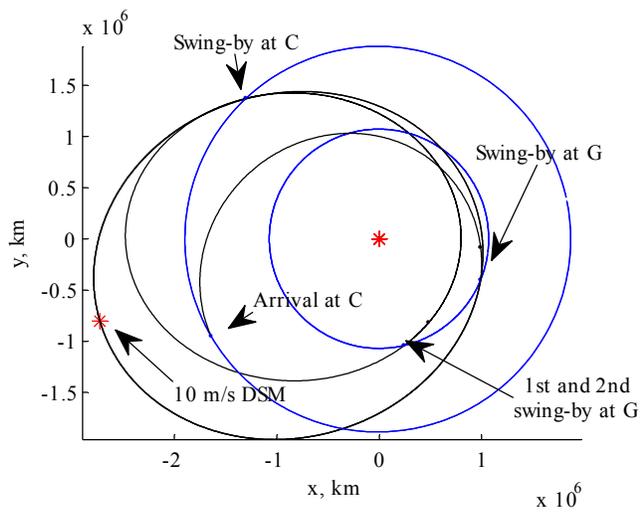

**Fig. 8: Best solution found by ACO-MGA, sequence GGCGC. The first leg is not plotted.**

Table 4: Characteristics of the ACO-MGA solution and the optimised 3D solution. The first leg is not considered.

| Variable | ACO-MGA | 3D optim. |
|---|---|---|
| $\Delta v_2$, m/s | 0 | 0 |
| $\Delta v_3$, m/s | 0 | 0 |
| $\Delta v_4$, m/s | 0 | 0 |
| $T_2$, d | 17.4 | 17.52 |
| $T_2$, d | 13.9 | 13.84 |
| $T_3$, d | 5 | 5.10 |
| $v_\infty$, km/s | 1.91 | 1.96 |

Table 5: Four solutions with low final relative velocity.

| | Sol. (a) | Sol. (b) | Sol. (c) | Sol. (d) |
|---|---|---|---|---|
| Sequence | GGCGC | GGCGC | GGCGC | GGCGC |
| $\Delta v_2$, m/s | 0 | 0 | 0 | 0 |
| $\Delta v_3$, m/s | -10 | 0 | 0 | 0 |
| $\Delta v_4$, m/s | 0 | -10 | -10 | 0 |
| $T_2$, d | 17.492 | 17.492 | 50.226 | 17.492 |
| $T_2$, d | 47.957 | 34.766 | 14.812 | 34.766 |
| $T_3$, d | 5.711 | 17.985 | 6.502 | 18.007 |
| $v_\infty$, km/s | 1.7097 | 1.7945 | 1.7972 | 1.8199 |

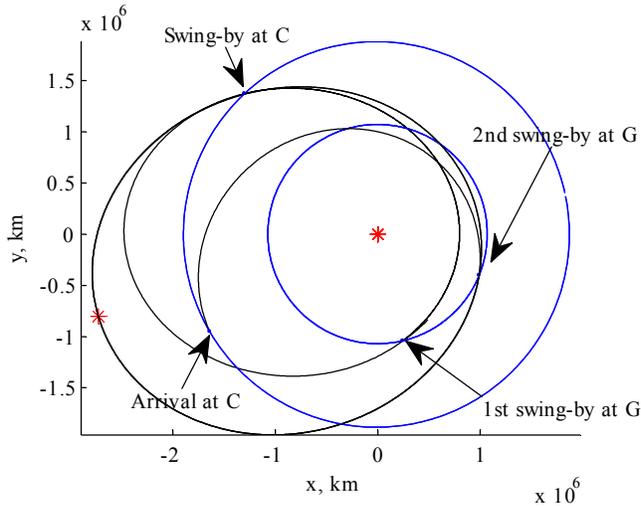

a)

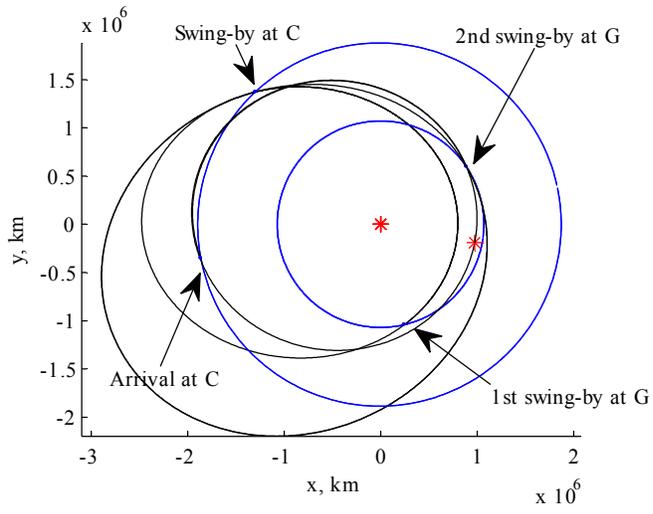

b)

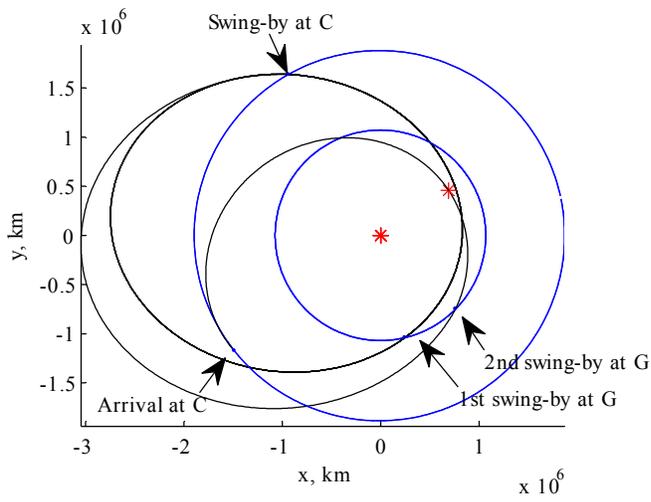

c)

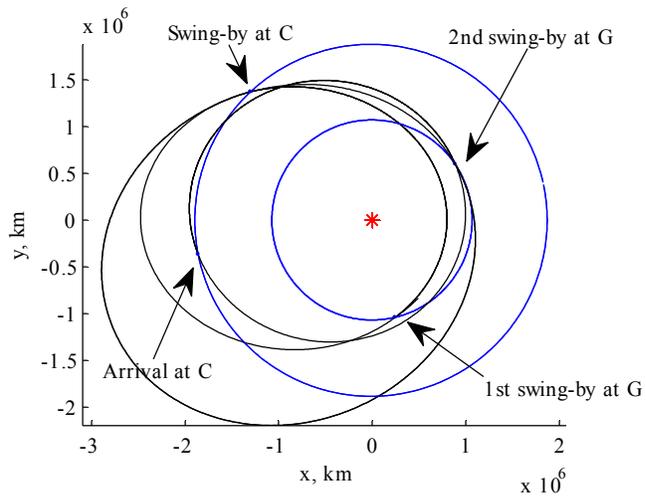

d)

**Fig. 9: Four solutions with low final relative velocity (see Table 5). The first leg is not plotted.**

**Table 6: Four solutions with short total time of flight.**

|  | Sol. (a) | Sol. (b) | Sol. (c) | Sol. (d) |
| --- | --- | --- | --- | --- |
| Sequence | GGCGC | GGGCC | GGGCC | GGCGC |
| $\Delta v_2$, m/s | 0 | -10 | -10 | -10 |
| $\Delta v_3$, m/s | -10 | -10 | -10 | -10 |
| $\Delta v_4$, m/s | 0 | -10 | -10 | 0 |
| $T_2$, d | 17.482 | 14.318 | 14.318 | 33.697 |
| $T_2$, d | 13.864 | 6.6194 | 6.6194 | 18.482 |
| $T_3$, d | 5.0154 | 30.889 | 46.362 | 15.339 |
| $v_\infty$, km/s | 1.9047 | 2.2068 | 2.2003 | 2.1437 |

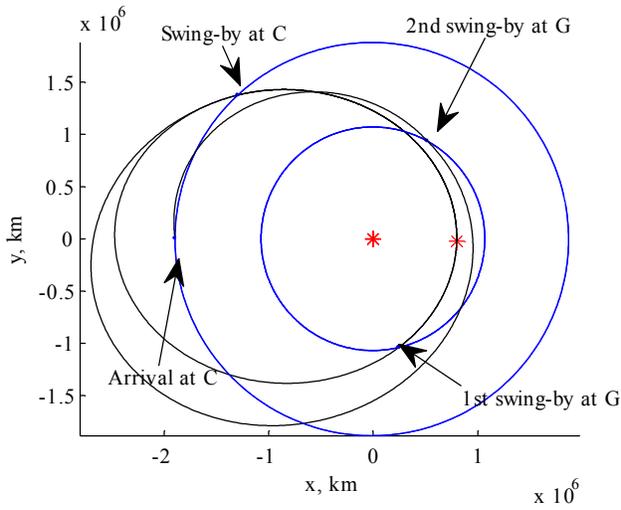

a)

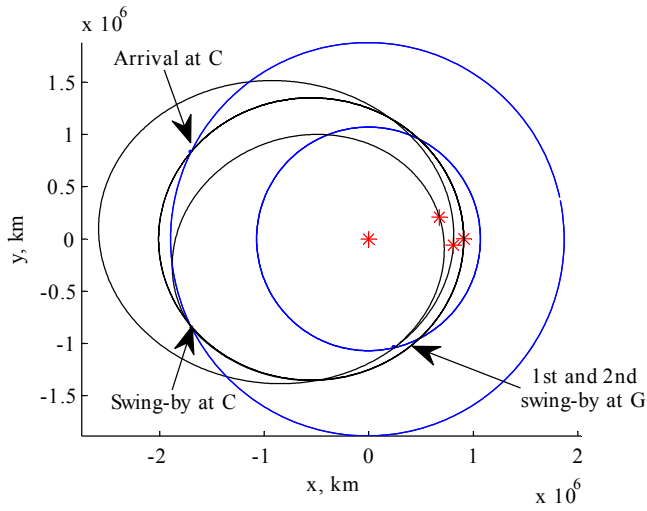

b)

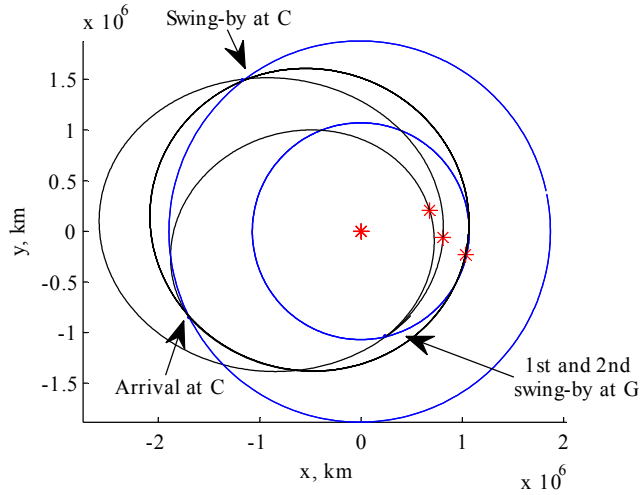

c)

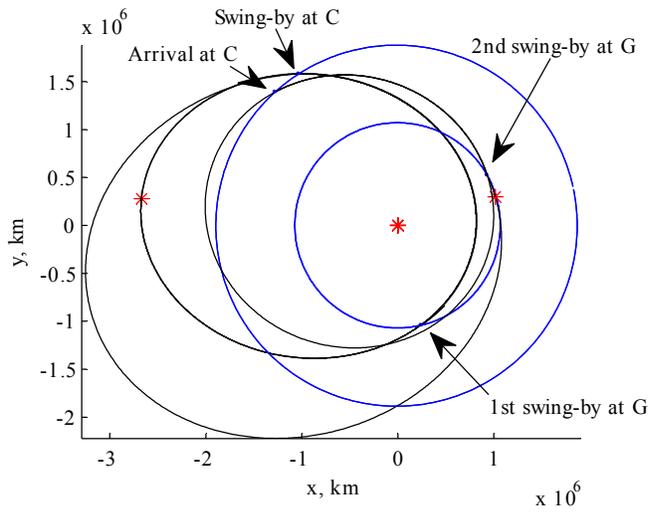

d)

**Fig. 10: Four solutions with short total time of flight (see Table 6). The first leg is not plotted.**

**Table 7: Best solution for each feasible sequence: objective value, final relative velocity, total DSM cost and total time of flight (considering only the three legs under design).**

| Sequence | $f_{obj} \equiv v_\infty$, km/s | $\Delta v$, m/s | $T$, y |
|---|---|---|---|
| GGCGC | 1.7097 | 10 | 71.1606 |
| GGGCC | 2.1940 | 10 | 70.0352 |
| GGGGC | 2.2354 | 30 | 67.9405 |
| GGCCC | 3.4084 | 30 | 68.3365 |

### 4.2 Cassini case study

Cassini is the ESA-NASA mission to Saturn. The planetary sequence designed for the mission, EVVEJS is particularly long, allowing a substantial saving of propellant.

Since the launch date is not taken into account in the optimisation, in the following test it is considered fixed. In a real mission design case, where the launch date is to be determined, the entire launch window can be discretised with a given time step, and a systematic scan of several dates within the whole launch window should be run. The launch direction $\alpha_0$ is also kept fixed in these tests, although it is easy to find heuristics for determining the value of this parameter, or discretise it and include it in the optimisation process as an additional variable.

For testing ACO-MGA we will make use of a 5-leg trajectory, with starting date $t_0 = -779$ d, MJD2000, corresponding to 13 November 1997. The following sets of parameters were used, to allow DSMs in the first 3 legs only:

$$m_{DSM}|_i \in \mathbf{q}_{1,i} = \{\pm 600, \pm 350, \pm 200, 0\} \text{ m/s}$$
$$n_{rev,1}|_i \in \mathbf{q}_{2,i} = \{0\}$$
$$n_{rev,2}|_i \in \mathbf{q}_{3,i} = \{0\} \qquad i = 1,2,3$$
$$f_{p/a}|_i \in \mathbf{q}_{4,i} = \{0,1\}$$
$$f_{1/2}|_i \in \mathbf{q}_{5,i} = \{0,1\}$$

$$m_{DSM}|_i \in \mathbf{q}_{1,i} = \{0\}$$
$$n_{rev,1}|_i \in \mathbf{q}_{2,i} = \emptyset$$
$$n_{rev,2}|_i \in \mathbf{q}_{3,i} = \{0\} \qquad i = 4,5$$
$$f_{p/a}|_i \in \mathbf{q}_{4,i} = \emptyset$$
$$f_{1/2}|_i \in \mathbf{q}_{5,i} = \{0,1\}$$

The planets available for swing-bys are $\mathbf{q}_{P,i} = \{\text{Venus, Earth, Jupiter}\}$, $i = 1,..,4$, while the target planet is obviously fixed to Saturn. The number of maximum full revolutions is fixed to 0, as it can be seen from the choice of parameters $n_{rev,1}$ and $n_{rev,2}$. This is done to limit the total time of flight of the mission. Since the trajectory is going outwards of the orbit of the Earth, every full revolution implies more than one additional year in the transfer time. The main aim of this case study, then, is to assess the ability of finding promising planetary sequences, using deep space manoeuvres. The total number of distinct solutions for this test is 7,112,448, and the average time to evaluate a solution is 1.26 ms. This translates in 8961.7 s (or about 2.5 hours) to systematically evaluate all the solutions.

The launch excess velocity module was bounded between 2 and 4 km/s. For the swing-bys of Earth and Venus, the radii of pericentre span from 1.1 to 5 $R_P$. A different choice is adopted for Jupiter. In fact, the mass of this planet is considerably bigger than the masses of Venus and Earth, so higher radii of pericentre are enough to achieve considerable deviations. It was decided to consider the range 5 to 100 $R_P$.

Regarding the choice of the objective function, it has to be noted that for all the missions to outer planets, the time of flight becomes very important, as very long missions are needed to reach farther destinations. Even limiting the number of complete revolutions to zero, is not enough to guarantee a mission with reasonable duration. Therefore, it is important to include the total time of flight $T$ in the objective function, in addition to the total $\Delta v$. Since the current algorithm cannot deal with multi-objective optimisation, the total time of flight and the $v_\infty$ are weighed inside the objective function, such that $f_{obj} = v_\infty + \sigma T$: for this test case the weight on $T$ was chosen to be $\sigma = 1/1000$ km/s/d.

The total time of flight has been limited to a maximum of 100 years: limiting the time of flight to lower values would over-constrain the search for optimal solutions. Instead, better results are obtained by allowing long solutions to be returned as feasible, and introducing their duration into the objective function.

GATBX and NSGA-II were run at first for 4000 and then for 6000 function evaluations and the same numbers of function evaluations were used as an upper limit on the function evaluations performed by ACO-MGA. The weights of ACO-MGA were set to $w_{planet}, w_{type} = 0$ for the first step, and $w_{planet}, w_{type} = 20 \cdot 7$ km/s for the second step. For each step, the number of iterations of ACO-MGA was set such that the expected total maximum number of function evaluations was 4000 for the first bunch of runs and 6000 for the second bunch of runs. In particular, for 4000 function evaluations, the number of iterations allocated to the first step was 200 and the number of iterations allocated to the second step was 300. For 6000 function evaluations, instead, the number of iterations allocated to the first step was increased to 600 and the number of iterations allocated to the second step was left equal to 300. With these settings, a run of ACO-MGA requires, on average, 1900 function evaluations and 161 s, if the upper limit is 4000, and 3300 function evaluations and 273 s, if the upper limit is 6000 evaluations. This is considerably faster than the exhaustive scan of the solution space and the number of function evaluations is lower than in the case of GATBX and NSGA-II, however the computational time for each single run of GATBX, which is fully coded in MATLAB®, is, on average, half of the one required to ACO-MGA as the access to the taboo and feasible lists is currently not optimal and coded in MATLAB®. Current developments have addressed this bottleneck and the results of the new implementation will be presented in future papers. NSGA-II is even faster as the code is fully in C. Therefore, a fair comparison would allow GATBX and NSGA-II to perform a higher number of function evaluations.

The parameters used for GATBX and NSGA-II are reported in Table 8. The comparative results for the two sets of runs are shown in Table 10. It can be seen that, for 1900 function evaluations, ACO-MGA found feasible solutions in 91% of the runs, compared to 25% of GATBX and 26% of NSGA-II. The average ACO-MGA solution is also slightly better than GATBX, and considerably better than NSGA-II. The performances of ACO-MGA increase significantly by using 3300 evaluations: all the runs produce a feasible solution, and in 80% of the cases, the best solution found is below 16 km/s. The average value of the solution also decreases to 15.434 km/s. It is interesting to note that, for GATBX, the average best solution found with 6000 evaluations is higher than for 4000: this is partly balanced by the fact that it finds feasible solutions in 28% of the runs. Another thing worth noticing is that NSGA-II finds more often feasible solutions than GATBX, but their quality is in average worse.

The best solution found by ACO-MGA (sequence EVVEJS) has an objective value of 6.9686 km/s: The characteristics of this solution can be found in Table 9, compared to the best solution found for the Earth-Saturn transfer problem (see http://www.esa.int/gsp/ACT/inf/op/globopt/edvdvdedjds.htm). The trajectory of the ACO-MGA solution is shown in Fig. 11 (a), while the 3D reference solution is in Fig. 11 (b).

It is worth to underline that the superior computing time of ACO-MGA is also due to the fact that for all runs, including the few that are not feasible, the algorithm finds partial solutions that are feasible up to the last transfer. As a consequence, the taboo lists of the last planetary encounter tend to grow significantly. A possible solution, on top of optimising the access to the list, would be to plan backward from the target planet in the case of transfers to the outer part of the solar system.

**Table 8. Parameters of GATBX and NSGA-II for the Cassini test case.**

| GATBX | | NSGA-II | |
|---|---|---|---|
| Parameter | Value | Parameter | Value |
| Common parameters | | | |
| *StallGenLimit* | +Inf | *pcross_bin* | 0.5 |
| | | *pmut_bin* | 0.5 |
| 4000 function evaluations | | | |
| *Generations* | 20 | *ngen* | 200 |
| *PopulationSize* | 200 | *popsize* | 20 |
| 6000 function evaluations | | | |
| *Generations* | 30 | *ngen* | 300 |
| *PopulationSize* | 200 | *popsize* | 20 |

Table 9. Characteristics of the ACO-MGA solution and the reference solutions.

| Variable | ACO-MGA | Reference |
|---|---|---|
| $v_0$, km/s | 3.14 | 3.259 |
| $\Delta v_1$, m/s | 600 | 480 |
| $\Delta v_2$, m/s | 350 | 398 |
| $\Delta v_3$, $\Delta v_4$, $\Delta v_5$, m/s | 0 | 0 |
| $v_\infty$, km/s | 4.21 | 4.246 |
| $T_1$, d | 168 | 167 |
| $T_2$, d | 423 | 424 |
| $T_3$, d | 53 | 53 |
| $T_4$, d | 596 | 589 |
| $T_5$, d | 2290 | 2200 |

Table 10. Comparison of the performances of ACO-MGA, GATBX, NSGA-II over 100 runs for the Cassini problem.

| Optimiser | Average best value, km/s | % runs < 16 km/s | % feasible runs |
|---|---|---|---|
| Max 4000 function evaluations | | | |
| ACO-MGA | 16.24 | 44% | 91% |
| GATBX | 16.349 | 14% | 25% |
| NSGA-II | 20.426 | 5% | 26% |
| Max 6000 function evaluations | | | |
| ACO-MGA | 15.434 | 80% | 100% |
| GATBX | 16.526 | 17% | 28% |
| NSGA-II | 20.122 | 7% | 37% |

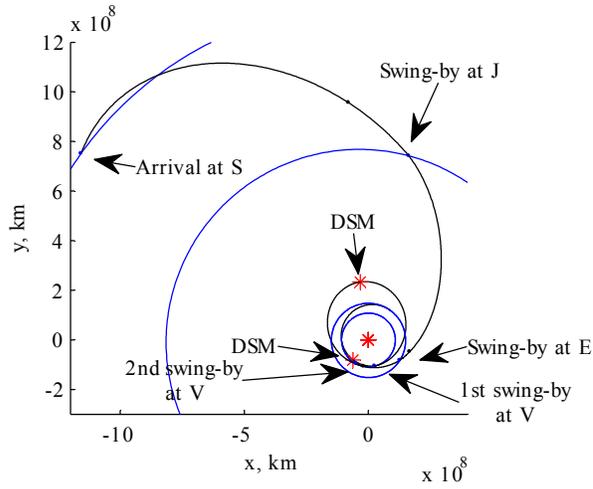

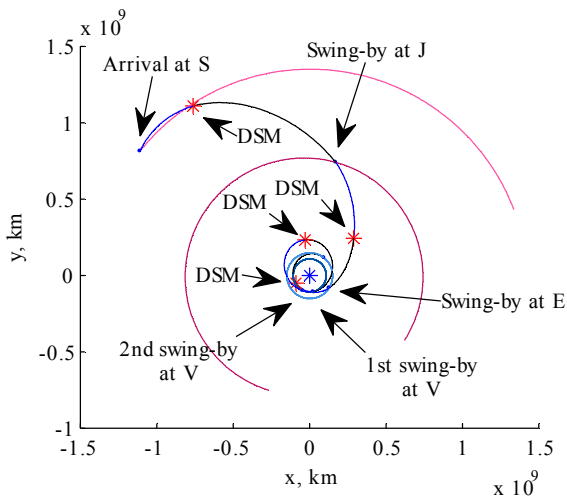

**Fig. 11: (a) ACO-MGA solution; (b) Cassini reference solution.**

It is interesting to sort the feasible sequences found by ACO-MGA according to the best objective value that they can achieve. Table 11 shows the objective value, the relative velocity at arrival, the total DSM cost and the time of flight of the best solution found for each sequence. Note that every solution has a trajectory associated to it, modelled as shown before, and thus taking into account the phasing problem. This means that these solutions could be re-optimised with a more detailed model (in particular including the third dimension), leading to actual transfer solutions.

However, differently from the previous test case, the alternative solutions have a considerably longer time of flight (> 13 years) and higher $v_\infty$, and therefore we consider them not interesting. Other promising solutions could possibly be found changing the definition of the problem, i.e. varying the launch angle and the sets $\mathbf{q}_{i,j}$.

Table 11: Best solution for each sequence: objective value, velocity at infinity, DSM cost and total time of flight.

| Sequence | $f_{obj}$, km/s | $v_\infty$, km/s | $\Delta v$, km/s | $T$, y |
|---|---|---|---|---|
| EVVEJS | 7.5832 | 4.1112 | 1.4 | 11.639 |
| EVVEES | 10.762 | 6.6954 | 1.4 | 27.978 |
| EVEEJS | 14.230 | 8.2067 | 1.15 | 18.624 |
| EVEEES | 16.135 | 10.283 | 1.4 | 32.188 |
| EVVVES | 21.261 | 10.437 | 1.2 | 33.038 |
| EVVVVS | 21.725 | 6.5036 | 1.55 | 13.791 |

**4.3  Launch date analysis**

As mentioned before, the algorithm, at the current state, does not perform any kind of search on the launch date $t_0$. In fact, this variable is not even included in the solution vector **s**. Rather, if the launch date is not fixed, but a launch window is available, a systematic scan can be performed to find the best launch date, and the corresponding solutions. This procedure is not always applicable: in fact, if re-running the algorithm for a small change in the launch date, the solutions that ACO-MGA finds are substantially different, then the systematic scan along $t_0$ is not feasible, and this variable must be included in the optimization process. If, on the other hand, a small displacement along $t_0$ causes a small change in the best solution found (e.g. same planetary sequence, possibly different types of transfers, similar objective value), then the systematic scan over $t_0$ is applicable to the identification of promising launch dates.

In order to verify this assertion, a test was run using the BepiColombo mission as a reference case [16]. BepiColombo is a multiple gravity assist mission to Mercury, currently under study at ESA and JAXA. In Ref. [16] an optimal transfer solution is provided, using two swing-bys of Venus to reach Mercury (sequence EVVMe). The optimal launch date is the 15 August 2013, i.e. $t_0^* = 4974.5$ d, MJD2000.

ACO-MGA was run, leaving the choice of the swing-by sequence and the other transfer parameters, like the number of revolutions, free. The objective was to minimise the relative velocity $v_\infty$ at Mercury.

Five different launch dates, in a window of 10 days around the one chosen by ESA, were considered, and for each one of them, 100 runs of ACO-MGA were performed. The best solutions for each launch date can be found in Table 12. Note that, the optimal launch date occurs 1 day before $t_0^*$, while earlier or later launch dates appear to be less convenient. In addition, all the solutions have the same planetary sequence.

The discrepancy between the value of $v_\infty$ found by ACO-MGA and the one in [16] has two reasons: the first is that ACO-MGA does not take into account the inclination of the planets, and the orbit of Mercury is highly inclined. The second is that the ESA solution was found as a part of a longer trajectory, and thus with a different objective.

The same reasons explain why, according to ACO-MGA, the ideal launch date is 1 day earlier. As a matter of fact, this is not a problem, and a subsequent local optimisation of the ACO-MGA solutions with a full model would tune the launch date.

This test demonstrates that the 2D model and the search process in ACO-MGA yield a distribution of solutions, along the departure time axis, that is consistent with the existing reference solution for this mission. The convergence of ACO-MGA is robust against variations of the launch date, as it consistently provides, solutions with unchanged planetary sequence that display a small variations in the cost function for small variations of the launch date. These properties together with the small computing time suggest that ACO-MGA can be used to systematically scan the launch dates in search for an optimal one. Note that, as the model in ACO-MGA is intended for the generation of first guess solutions, the scan of the launch dates is expected to provide only an estimated location of the optimal point along the launch date coordinate.

**Table 12: Best solutions to Mercury found by ACO-MGA for different launch dates.**

| Launch date $t_0$ | Optimal sequence | $f_{obj} \equiv v_\infty$, km/s |
|---|---|---|
| $t_0^* - 5\,\text{d}$ | EVVMe | 5.98 |
| $t_0^* - 1\,\text{d}$ | EVVMe | 5.84 |
| $t_0^*$ | EVVMe | 6.10 |
| $t_0^* + 1\,\text{d}$ | EVVMe | 6.62 |
| $t_0^* + 5\,\text{d}$ | EVVMe | 6.72 |

## 5 CONCLUSION

The paper introduced a novel formulation of the automatic complete trajectory planning problem and proposed a new algorithm (ACO-MGA), based on the ant colony paradigm, to generate optimal solutions to this problem. Each solution is a complete, scheduled plan. A specific trajectory model was developed to efficiently generate families of scheduled trajectories for multi-gravity assist transfers, once a plan is available. The 2D trajectory model proved to be accurate enough to closely reproduce known MGA transfers even with moderate inclinations. Furthermore, the scheduling of the trajectories is fast and reliable allowing for the evaluations of thousands of plans in a short time.

ACO-MGA operates an effective search in the finite space of possible plans. The algorithm demonstrated a remarkable ability to find good solutions with a very high success rate, outperforming known implementations of genetic algorithms. As ACO-MGA requires very little information on the MGA problem under investigation, it represents a valuable tool for the complete automatic design of future space missions. Future work aims at a more efficient handling of the lists, which is currently the major bottleneck of the ACO-MGA implementation.

## ACKNOWLEDGEMENTS

This study was completed when Matteo Ceriotti was a Ph.D. research student in the Space Advanced Research Team at the Department of Aerospace Engineering of the University of Glasgow.